\documentclass[showpacs,amsmath,amssymb,prl,twocolumn]{revtex4-1}

\usepackage{graphicx}
\usepackage{dcolumn}
\usepackage{bm}

\begin{document}

\preprint{APS/123-QED}

\title{Probing single-photon state tomography using phase-randomized coherent states}

\author{P. Valente, H. Failache and A. Lezama}
\email{alezama@fing.edu.uy}
\affiliation{Instituto de F\'{\i}sica, Facultad de Ingenier\'{\i}a,
Universidad de la Rep\'{u}blica,\\ J. Herrera y Reissig 565, 11300
Montevideo, Uruguay}%

\date{\today}

\begin{abstract}
Quantum processes involving single-photon states are of broad interest in particular for quantum communication. Extending to continuous values a recent proposal by Yuan et al \cite{YUAN16}, we show that  single-photon quantum processes can be characterized using phase randomized coherent states (PRCS) as inputs. As a proof of principle, we present the experimental investigation of single-photon tomography using PRCS. The probability distribution of field quadratures measurements for single-photon states can be accurately derived from the PRCS data. As a consequence, the Wigner function and the density matrix  of single-photon states are reconstructed with good precision. The sensitivity of the reconstruction to experimental errors and the number of PRCS used is addressed. 
\end{abstract}

\maketitle

\section{Introduction}
A quantum process (QP) can be thought as a "black box" whose input consists of a quantum system in a given state. The state is modified during the process and measurements are performed at its output. Frequently, QPs are implemented with light so that the quantum state of the system refers to the  state of a given light mode. The characterization of an unknown QP is generally a complex task that requires the probing of the QP with different light states forming a basis of the Hilbert space \cite{RAHIMI11}. However, the full characterization of a QP is not always necessary. An approximate characterization can be achieved through the response of the QP to a finite number of selected input states \cite{REHACEK10}. Also in many cases, only the response of the QP to an input consisting of light in a single-photon Fock state is of interest. Examples of this are quantum key distribution (QKD) protocols \cite{BENNETT84,EKERT91,SCHMITT07}. In these processes, Alice prepares the input state encoding information in the photon degrees of freedom (polarization, time-frequency, orbital angular momentum). The physical transmission line, Bob's receiving apparatus and a possible intervention of an eavesdropper, Eve, constitute the QP. The characterization of such process is of great importance to its security since it may reveal the presence of the eavesdropper. 

QKD distribution protocols require the use of single-photon states. However, deterministic single-photon sources are still not readily available \cite{LOUNIS05,HE13} and probabilistic single-photon sources, such as heralded parametric down conversion, are inefficient and too complex and expensive to be practical in commercial devices. In consequence, many practical implementations of quantum information protocols use, instead of single-photon pulses, attenuated coherent state pulses that can be produced on demand. However, the use of weak coherent state (WCS) pulses carries its disadvantages. On one hand it is inefficient since most pulses carry no photon and, more importantly, it jeopardizes security since the probability for two or more photons is non-zero which allows for different forms of attack by an eavesdropper \cite{BENNETT92,DUSEK00,BRASSARD00,HWANG03,GOTTESMAN04}.

The security issue of QKD with WCS has been solved through the decoy state method \cite{LO05,MA05,WANG05,ZHAO06,INAMORI07,SCHMITT07,SUN13,LIM14,WANG14}. In essence, Alice, the sender, uses a random set of WCS of different amplitudes. On the receiving end, Bob analyses the received data and is able, after communicating with Alice, to detect the presence of a possible eavesdropper. More specifically, Bob and Alice are able to extract from the information obtained with WCS, the yield through the communication channel of the single-photon states and compute the amount of information leaked to a possible eavesdropper \cite{LO05,LIM14}. \\

Extending the idea of the decoy state method, Yuan et al. have shown in a recent theoretical article \cite{YUAN16} that phase randomized coherent states (PRCS) can be used for the computation of the single photon-yield through an arbitrary QP. Their analysis was restricted to processes ending with a binary measurement outcome (``click" or ``no-click").

In this article we present a generalization of the method in \cite{YUAN16} to QPs ending up with a continuous variable measurement outcome. We show that PRCS can be used to accurately determine the probability distribution for continuous variable measurement carried upon single-photon states. To illustrate our result, we present a proof of principle experiment in which  PRCS are used to probe the QP used in quantum tomography \cite{LVOVSKY01}. As a result, the quantum tomography  of a single-photon state was efficiently simulated using a small ensemble of PRCS \cite{Comment}.\\ 

We consider temporal optical field modes of constant amplitude having a finite duration $T$. A stationary light beam can thus be viewed as many consecutive realizations of the mode state. More precisely, a well stabilized laser beam can be considered to a good approximation as the multiple realization of a coherent state $\vert \alpha\rangle$ of the temporal mode. If the phase of the laser beam is varied at random between different realizations of the temporal mode, then the prepared state is a PRCS with mean photon-number $\mu=\vert \alpha\vert^{2}$.

Following the usual quantum tomography sequence \cite{LVOVSKY09}, we have initially determined an approximation to the field quadrature probability density (marginal distribution) of the single photon state. For this, the marginal distributions of several PRCS with different mean photon-numbers were acquired and combined using the procedure described below to derive an approximation to the marginal distribution of the single-photon state. In a second time, the approximate Wigner function of the single-photon state and the corresponding density matrix were derived. Our example shows that excellent approximations are obtained even for a reduced number of mean photon-numbers values.

\section{Theoretical background}\label{background} 

The density matrix for a phase averaged coherent state of mean photon number $\mu$ is given by:

\begin{eqnarray} \label{mat_densidad} 
\rho_{\mu}=\sum_{k=0}^{\infty} P_{\mu}(k)\vert k \rangle \langle  k\vert
\end{eqnarray}
where $\vert k \rangle$ designates a Fock state with photon number $k$ and
\begin{eqnarray} \label{Pk} 
P_{\mu}(k)=\frac{\mu^{k}}{k!}e^{-\mu}
\end{eqnarray}

If a measurement is performed on the system in the state $\rho_{\mu}$, the probability $X_{\mu}$ for an outcome $x$ is given by:
\begin{eqnarray} \label{X} 
X_{\mu}(x)=\sum_{k=0}^{\infty} P_{\mu}(k)P(x\vert k)
\end{eqnarray}
where $P(x\vert k)$ denotes the probability for an outcome $x$ when the system is prepared in the Fock state $\vert k \rangle$.\\

Unlike in \cite{YUAN16} where expression (\ref{X}) was applied to a binary measurement with two discrete outcomes (``click" or ``no-click"), here we consider a measurement with a continuous variable outcome $x$. In consequence, $X_{\mu}(x)$ is the probability density for the measurement outcome to be in the range $[x,x+dx]$. Notice that 
\begin{subequations}\label{Xmu}
\begin{eqnarray}  
X_{\mu}(x), P(x\vert k)\geq 0\\ \label{positivos}
\int X_{\mu}(x)dx= \int P(x\vert k)dx =1
\end{eqnarray}
\end{subequations}

Following \cite{YUAN16}, we introduce the notation $Y_{k}\equiv P(x\vert k)$. The values of $\mu$ and $X_{\mu}$ (we drop the $x$ dependence for compactness) can be experimentally determined. Since $X_\mu$ is linearly related to the $Y_k$'s via Eq. (\ref{X}), the knowledge of $X_\mu$ for a sufficiently large ensemble of values of $\mu$ should allow the approximate determination of the single-photon yield $Y_k$. Extending the decoy state method \cite{LO05}, Yuan et al have derived several estimates for $Y_1$ depending on the available values of the mean-photon-numbers $\mu$ \cite{YUAN16}. Their results concerning the estimate of $Y_1$ remain valid here keeping in mind the $Y_{k}$ and $X_{\mu}$ refer now to continuous variable distributions.\\ 

If only the vacuum and a single value of $\mu$ are available, the estimate for $Y_1$ is:
\begin{eqnarray} \label{Y1} 
Y_1^{est(1)}=\frac{(X_\mu e^{\mu}-X_0)}{\mu}
\end{eqnarray}
which is an upper bound of $Y_1$ (see Eq. (15) in \cite{YUAN16}).

On the other hand, when the information regarding the vacuum state ($\mu=0$) and two or more values of $\mu\neq 0$ is available the estimate of $Y_1$ is given by:
\begin{eqnarray} 
Y_1^{est(L)}=\sum_{j=0}^{L}\lambda_j X_{\mu_j}  \label{YL}\\
\lambda_0=-\mu_1\mu_2\cdots\mu_L\sum_{j=1}^{L}\frac{\mu_j^{-2}  }{\prod_{1\leq n \leq L,n\neq j}\left( \mu_n - \mu_j\right) },\nonumber \\
\lambda_{j\geq 1}=\mu_1\mu_2\cdots\mu_L\frac{\mu_j^{-2}e^{\mu_j} }{\prod_{1\leq n \leq L,n\neq j}\left( \mu_n - \mu_j\right)}.\nonumber
\end{eqnarray}

This estimate is an upper(lower) bound for $Y_1$ depending on $L$ being odd(even) [see Eq. (19) in \cite{YUAN16}].\\

We have  used expressions (\ref{Y1}) and (\ref{YL}) for the determination of the probability density of the field quadrature $\hat{X}_{\theta}\equiv (ae^{-i\theta}+a^{\dagger}e^{i\theta})/2$ for the single photon state $\vert 1\rangle$. $a$ and $a^{\dagger}$ are the usual photon annihilation and creation operators. Since all states considered in this work have no definite phase, the probability distribution of $X_{\theta}$ is independent of $\theta$, we will consequently drop the suffix. Notice that with the above definition the variance of $\hat{X}$ in the vacuum state (or in a coherent state) is $\langle 0\vert \hat{X}^{2}\vert 0\rangle=1/4$.

The expected probability distribution for the outcomes of $\hat{X}$ for a phase averaged coherent state of mean photon-number $\mu$ is given by (\ref{X}) where $P(x\vert k)$ are the squares of the well known Hermite-Gauss eigenfunctions of the harmonic oscillator with $k$ excitations.

\section{Experiment}

\begin{figure}[h]
\centering
\fbox{\includegraphics[width=\linewidth]{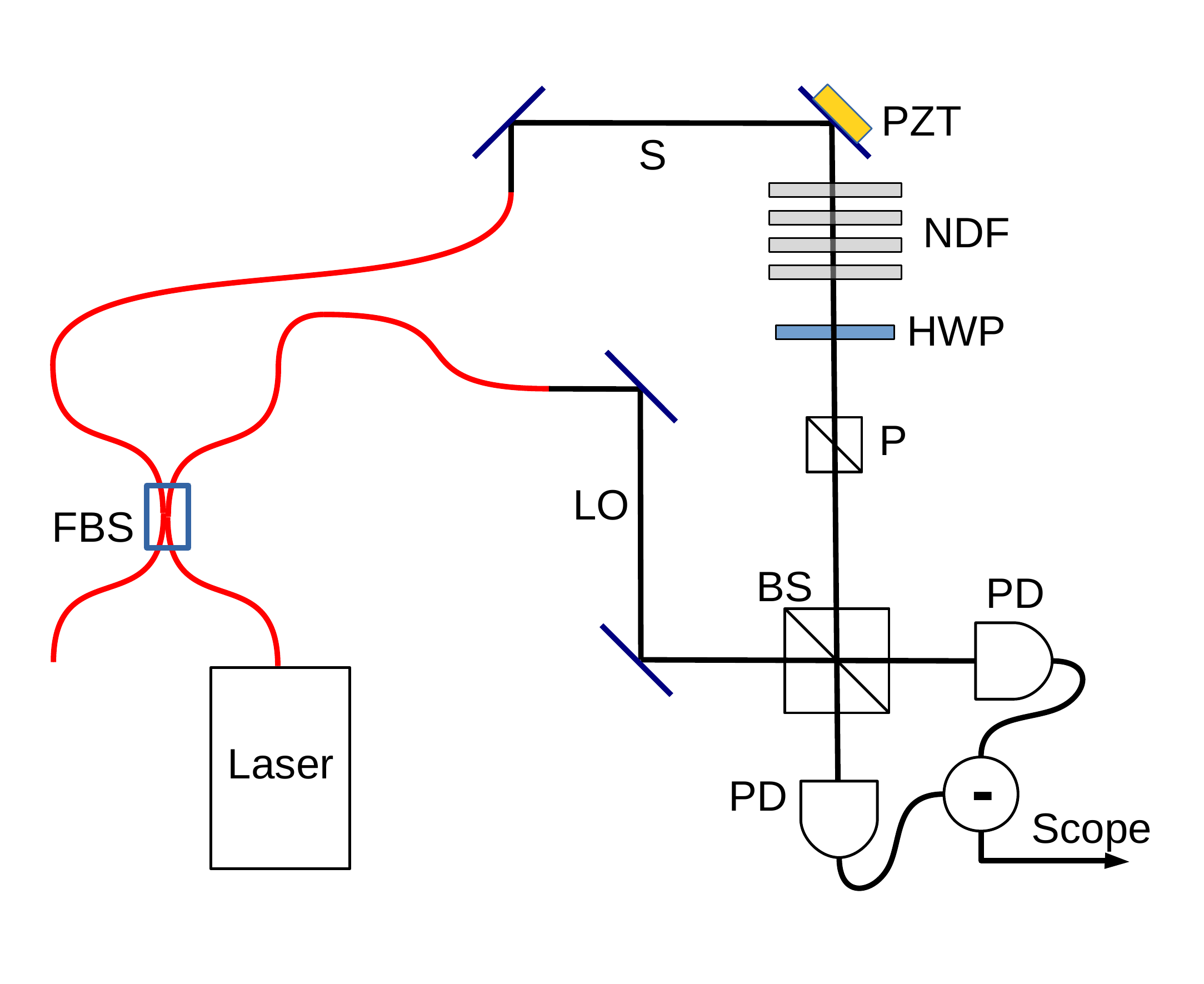}}
\caption{(Color online) Experimental setup. FBS: fiber beam splitter, BS: beam splitter, PZT: piezo-electric actuator, NDF: neutral density filters, P: polarizer, HWP: half-wave plate, PD: photodetectors, S: signal beam path, LO: local oscillator beam path.}
\label{setup}
\end{figure}

The  experimental setup is presented in Fig. \ref{setup}. A single-mode extended cavity diode laser operating around 795 nm in cw regime was used. The signal and local oscillator (LO) light beams were obtained from the laser output beam with the help of an optical fiber beam splitter (FBS). The power in the local oscillator was approximately 10 mW. The signal beam power after the FBS was attenuated to picowatt values using a combination of neutral density filters, half-wave plate and polarizer. The signal beam and the LO are recombined in a beamsplitter in the usual balanced homodyne detection configuration. Detection is achieved with a photodetector pair specifically designed for balanced detection (Thorlabs PDB120A).

The relative phase difference between the signal and the LO optical fields was varied with a mirror mounted on piezo-electric actuator (PZT). Using equal intensities for the signal and LO a contrast better than 98\% was typically observed in the interference after the recombining beamsplitter.  During data acquisition, in order to produce the phase randomization, a triangular voltage ramp is applied to the PZT with an amplitude corresponding to an approximately integer number of $2\pi$ phase differences and a frequency near 100 Hz. With this choice, during data acquisition,  the relative phase between signal and LO effectively explores randomly the $0$ - $2\pi$ interval. 

The difference of the two photodetector outputs, which is proportional the signal field quadrature, was monitored in a digital oscilloscope (Tektronix DPO4102B, 1 GHz bandwidth, maximum sampling rate: 5 GS/s). 

The balanced detector signal was sampled with a sampling frequency of 2.5 MHz. The recorded sample value corresponds to the average over the sampling interval of many scope internal acquisitions taken at the fastest scope sampling rate (High Resolution mode). This ensures that the amplitude of the sampled signal fluctuations scales as the inverse square-root of the interval duration. Each sampling record consisted of $10^{7}$ samples. For each record, the mean value of the fluctuations was subtracted and a histogram of the fluctuations computed. The final histogram was the average over 100 similar sampling records.  

With the signal beam blocked, we checked that the variance of the quadrature fluctuations increases linearly with the LO power as expected for vacuum fluctuations. 

\section{Results and discussion}
\subsection{Mean photon number determination}

After acquisition, the experimental histograms were rescaled. From the histogram corresponding to the vacuum state, we computed the sample variance of the fluctuations. This can be achieved with high relative precision (typically $10^{-4}$), given the large number of acquired data points. Subsequently, the horizontal axis was rescaled for the vacuum state fluctuations to have a variance equal to $1/4$. The vertical axis scale of each histogram was chosen such that the sum of all the histogram columns multiplied by $\Delta x$ equals 1 (normalization condition),  with $\Delta x$ being the renormalized distance between two consecutive histogram columns.

\begin{figure}[h]
\centering
\fbox{\includegraphics[width=\linewidth]{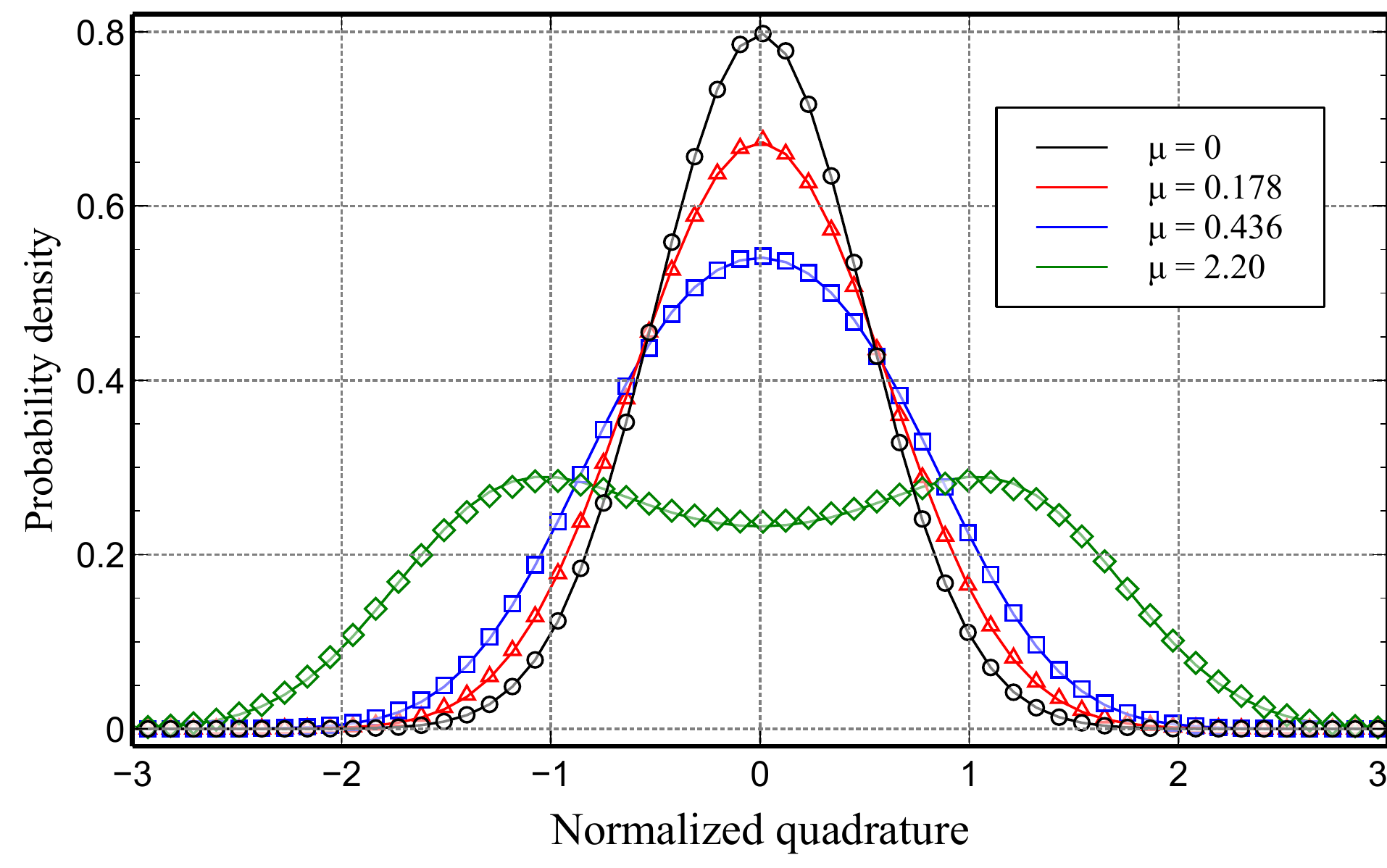}}
\caption{(Color online) Symbols: Experimental histograms for different mean photon-number values. Solid lines: Calculated $X_{\mu}(x)$ from Eq. (\ref{X}) using the values of $\mu$ indicated in the figure.}
\label{histogramas}
\end{figure}

The histograms obtained for the vacuum state and three different signal field intensities are presented in Fig.(\ref{histogramas}). Notice that the histogram corresponding to the vacuum state is well fitted by a Gaussian distribution. The histograms corresponding to non-zero signal field intensities were fitted to Eq. (\ref{X}) using $\mu$ as an adjustable parameter \cite{MUNROE95}. The best fits were obtained for $\mu= 0.178\pm 0.002$, $0.436\pm 0.004$ and $2.20\pm 0.01$. The corresponding plots of $X_{\mu}(x)$ are represented as continuous lines in Fig. (\ref{histogramas}). Notice the good agreement between the experimental data and the theoretical prediction. While very small, there are some deviations between the recorded histograms and the theoretical distributions given by Eq. (\ref{X}). These deviations are mainly due to imperfect phase randomization and electronic noise background.

We take the values of $\mu$ arising from the fitting procedure described above as the experimental mean photon-number values. We have checked that these values are consistent with the photon-number estimation obtained from the light power at the optical fiber beam-splitter output and the calibration of the neutral density filters used for attenuation. We have also verified that the mean photon numbers scale linearly with the signal light power and the duration of the sampling interval. 

\subsection{Estimated marginal distributions}

\begin{figure}[h]
\centering
\fbox{\includegraphics[width=\linewidth]{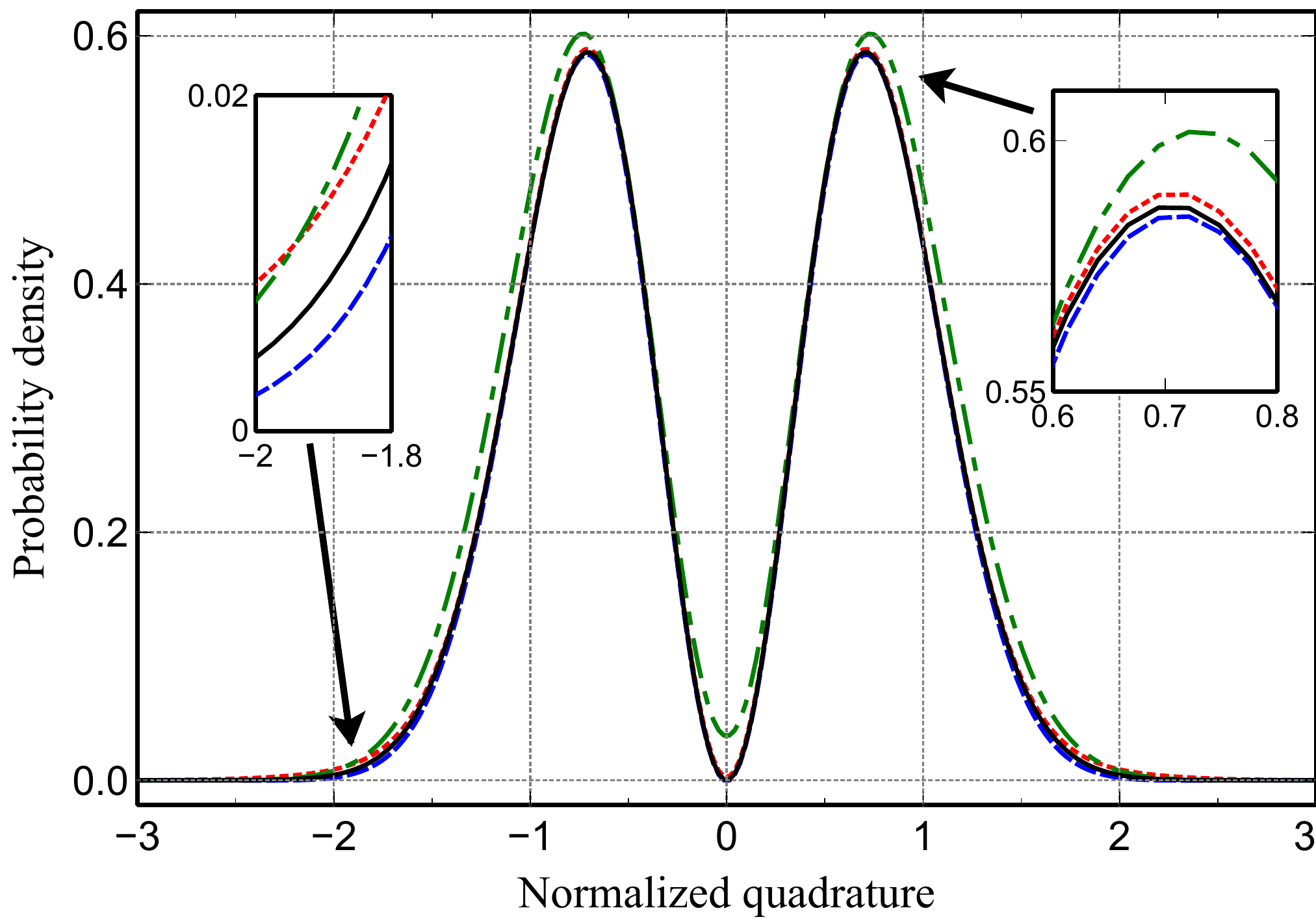}}
\caption{(Color online) Single photon quadrature probability density distribution. Solid line: exact. Dash-dotted: Estimation according to Eq. (\ref{Y1}) using $\mu=0.178$. Dashed: Estimation according to Eq. (\ref{YL}) using $\mu=0.178, 0.436$. Dotted: Estimation according to Eq. (\ref{YL}) using $\mu=0.178, 0.436,2.20$. For all the estimations de theoretical values of $X_{\mu}(x)$ given by Eq. (\ref{X}) were used.}
\label{teo}
\end{figure}

Figure \ref{teo} shows the estimates of the single photon quadrature probability distribution obtained with Eq. (\ref{Y1}) and (\ref{YL}) from the \emph{theoretical} marginal distributions corresponding to the observed mean-photon numbers. The dashed-dotted line corresponds to the use  of Eq. (\ref{Y1}) to a single non-zero $\mu$ in addition to the vacuum state. The dashed and dotted lines correspond to Eq. (\ref{YL}) applied to the vacuum and two and three non-zero $\mu$'s respectively. As predicted in \cite{YUAN16} the estimation for a single non-zero PRCS is an overestimation of the exact probability distribution. Closer approximations are obtained when the number of PRCS's used is increased. Notice that tight upper and lower bounds to single-photon probability distribution are obtained if numbers of $\mu$'s with different parities are used.

It is worth reminding that the distribution obtained from the application of Eqs. (\ref{Y1}) and (\ref{YL}) are not assured to satisfy the requirements in Eqs. (\ref{Xmu}). These requirements are only met approximately \cite{YUAN16}.\\

\begin{figure}[h]
\centering
\fbox{\includegraphics[width=\linewidth]{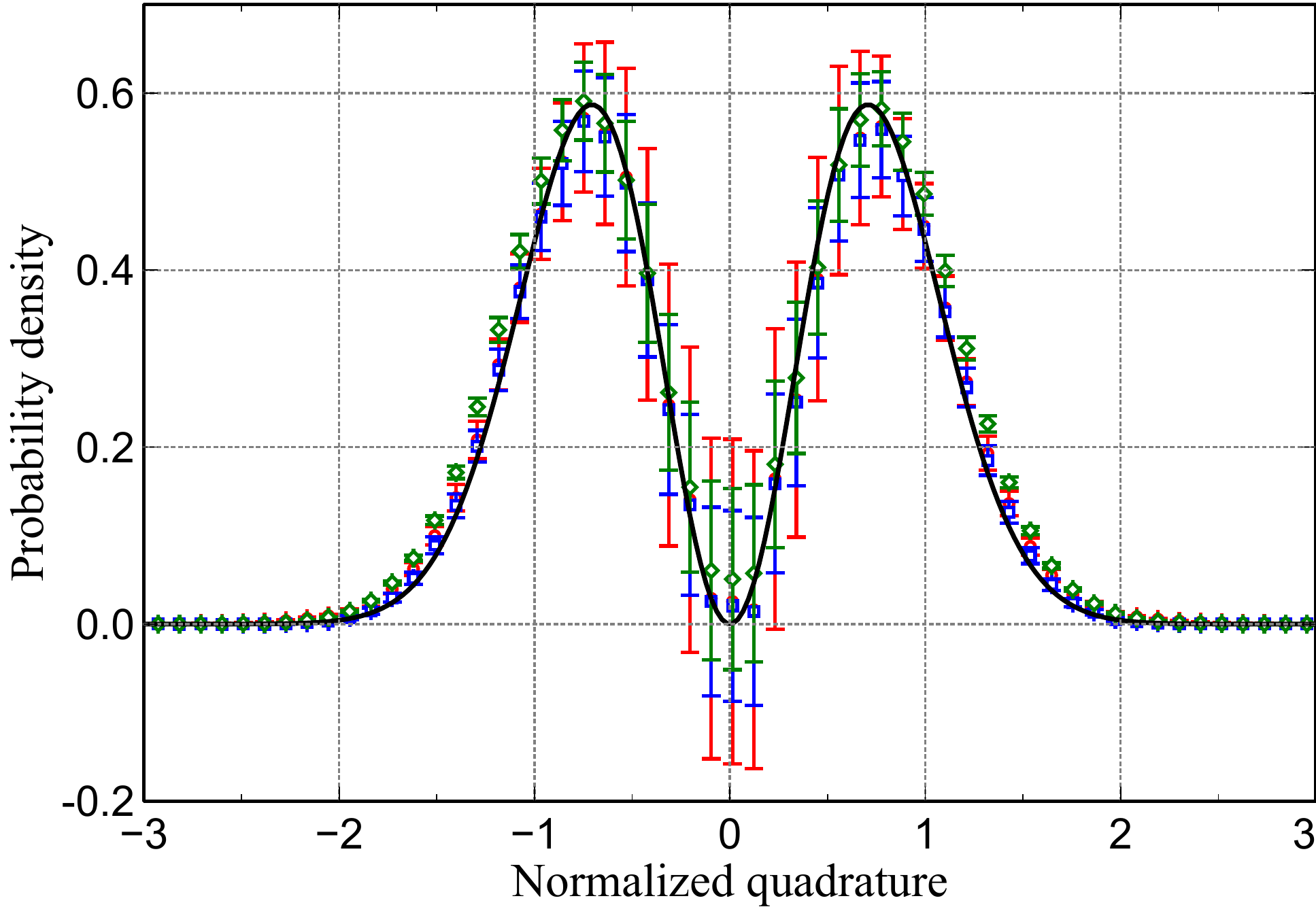}}
\caption{(Color online) Estimates of the single photon quadrature probability density using the experimental histograms in Fig. (\ref{teo}).  Diamonds (green): Estimate using Eq. (\ref{Y1}) and $\mu=0.178$. Squares (blue): Estimate using Eq. (\ref{YL}) and $\mu=0.178, 0.436$. Circles (red): Estimate using Eq. (\ref{YL}) and $\mu=0.178, 0.436,2.20$. Solid line: theoretical value.}
\label{estimates}
\end{figure}

Figure \ref{estimates} shows the estimated marginal distributions for the single-photon state computed directly from the experimental histograms of the quadrature fluctuations of the PRCS. As expected, the single-photon marginal distribution is affected by measurement uncertainties. The error bars in Fig. \ref{estimates} were computed from Eqs. (\ref{Y1}) and (\ref{YL}) using standard error propagation formula. While the statistical uncertainties in the histograms play a relatively minor role (in view of the large number of acquired samples) the error bars are mainly determined by the uncertainties in the values of the mean-photon numbers. The largest errors occur near the center of the distribution where the vacuum fluctuations and the fluctuations corresponding to small mean photon numbers are maximum and add-up. Notice the relatively large errorbars in spite of the small ($\sim 10^{-2}$) relative uncertainties of mean-photon numbers. The negative values included within error bars at the center of Fig. \ref{estimates} dramatically illustrate the well known fact that experimental errors can result in non-physical values in quantum-state reconstruction \cite{LVOVSKY09} .

\subsection{Wigner function and density matrix reconstruction}

In quantum tomography \cite{LVOVSKY09} the quantum state Wigner function can be derived from the observed marginal distributions using the inverse Radon transformation. However, because the marginal distribution are affected by experimental error,  this method  does not ensure that the obtained Wigner function corresponds to a physical state. To circumvent this problem, other techniques such as maximum likelihood or maximum entropy \cite{LVOVSKY09} have been suggested. These techniques are suitable for the reconstruction of the density matrix in the present case. However, based on the good agreement between experiment and theoretical expectations for PRCS observed in Fig. (\ref{histogramas}), we follow a simpler procedure only based in the experimental determination of the PRCS mean-photon numbers. We stress that as a consequence of this simplification, the  physical constraints on the Wigner function and the reconstructed density matrix are relaxed as discussed below. 

Since the estimated marginal distribution is a linear combination of the observed marginal distributions for PRCS [Eq. (\ref{YL})], the corresponding reconstructed Wigner function $W^{est}(r)$ must be linearly related to PRCS's Wigner functions $W_{\mu_{j}}(r)$ [$r$ is the radial distance to the phase-space origin].
\begin{eqnarray}
W^{est}(r)=\sum_{j=0}^{L}\lambda_j W_{\mu_{j}}(r) \label{reconstruyo}
\end{eqnarray}

\begin{figure}[h]
\centering
\fbox{\includegraphics[width=\linewidth]{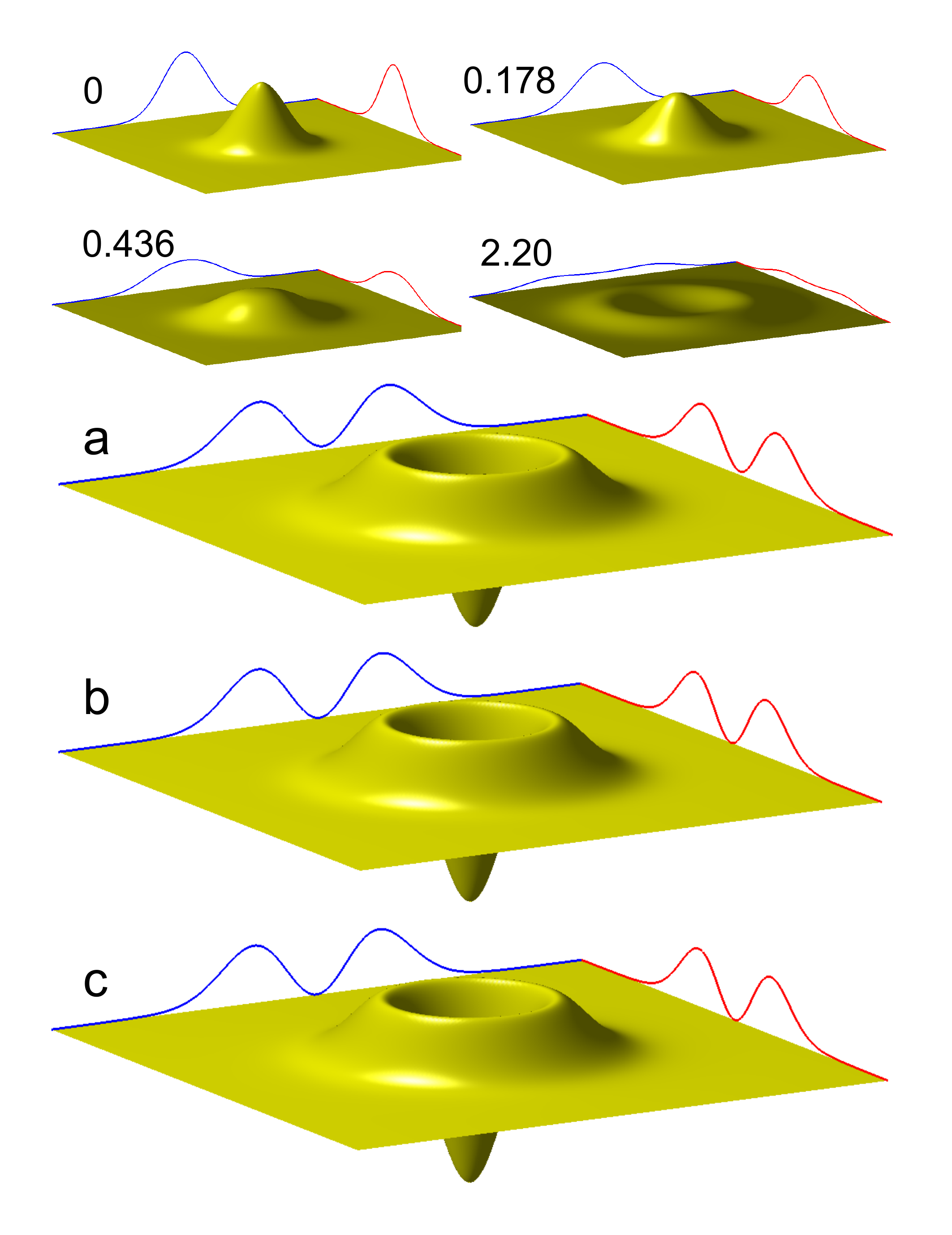}}
\caption{(Color online) Two upper rows: Calculated Wigner functions and marginal distributions for phase-averaged coherent states with $\mu=0$, $0.178$, $0.436$ and $2.20$ respectively. Three lower rows: Reconstructed Wigner function and marginal distributions for the single photon state using Eq.  (\ref{reconstruyo}) with different number of non-zero mean photon numbers. a) Vacuum plus one. b) Vacuum plus two. c) Vacuum plus three. [$\mu$ values indicated in first column of Table \ref{tabla}]}
\label{embudo}
\end{figure}

Figure \ref{embudo} shows the theoretical Wigner functions plots $W_{\mu_{j}}(r)$ for $\mu=0$, $0.178$, $0.436$ and $2.20$ and the corresponding reconstructed Wigner functions $W^{est}(r)$ obtained using Eq. (\ref{reconstruyo}) for an increasing number of non-zero PRCS (first column in Table \ref{tabla}). Even for a single value of $\mu\neq 0$, the similarity to the single-photon Wigner function is striking (see \cite{GERRY05}). Better approximations to the exact Wigner function are obtained as the number of PRCS increases. \\

In a similar way, the estimated density matrix $\rho^{est}$ (in the Fock states basis) can be calculated as a linear combination of the density matrices $\rho_\mu$ corresponding to the phase-averaged coherent states. Notice that by this means $\rho^{est}$, albeit Hermitian, is not assured to posses the usual physical requirements ($\rho >0$ and $Tr(\rho)=1$) which are only approximately satisfied. 

\begin{table}[h]
\centering
\begin{tabular}{|l|l|l|l|}
\hline
$\mu \neq 0$ & $Tr(\rho^{est})$ & $\vert \rho^{est}-\vert 1\rangle\langle 1\vert \,\vert$ & Eigenvalues $< 0$ \\ \hline
$0.178$              & $1.095$      & $8.9\times 10^{-2} $          & No                   \\ \hline
$0.178$, $0.436$            & $0.985$      & $1.3\times 10^{-2} $        & Yes                  \\ \hline
$0.178$, $0.436$, $2.20$         & $1.013$      & $8.3\times 10^{-3} $  & No      \\ \hline            
\end{tabular}
\caption{\label{tabla}Reconstructed density matrix characteristics for different number of phase-averaged coherent states}
\end{table}

A common practice to evaluate the quality of state reconstruction is to compute the fidelity and the purity for the reconstructed density matrix. However, since our procedure does not enforce the normalization and positivity of the reconstructed density matrix, the fidelity and purity are ill-defined. Instead, in order to quantitatively characterize the quality of the reconstructed density matrix we use the distance $\vert \rho^{est}-\vert 1\rangle\langle 1\vert \,\vert$ between $\rho^{est}$ and the target single photon density matrix  [we use $\vert M \vert\equiv Tr(MM^{\dagger})^{1/2}$]. For reference: the distance between two different Fock states is $\sqrt{2}$.

Table \ref{tabla} summarizes the characteristics of the reconstructed density matrices for different numbers of phase-randomized coherent states in addition to the vacuum. Notice that as the number of non-zero intensity PRCS increases, $Tr(\rho^{est})$ closely approaches the physical requirement of unity and the distance to the exact single photon density matrix decreases. Also when the number of non-zero intensity PRCS is even, $\rho^{est}$ posses non-physical  negative eigenvalues. As clearly shown in Fig. \ref{embudo} and table \ref{tabla}, most of the information required for the single-photon state reconstruction originates from the vacuum and the smallest mean photon number. Additional PRCS with increasing photon-numbers only provide small improvements to the single-photon density matrix reconstruction.

\section{Conclusions}

We have shown experimentally that PRCS can be simply and efficiently used to probe a tomography QP for single-photon temporal states. Using a small ensemble of PRCS the probability distribution of the field quadrature was estimated and the single photon Wigner function and density matrix were reconstructed with good precision. Similar techniques can be applied to characterize and simulate any single photon QP without the need for actual single-photon sources.\\

A.L. is thankful to D. Gauthier for motivating discussions. This work was supported by ANII, CSIC and PEDECIBA (Uruguayan agencies).\\


\begin{thebibliography}{26}%
\makeatletter
\providecommand \@ifxundefined [1]{%
 \@ifx{#1\undefined}
}%
\providecommand \@ifnum [1]{%
 \ifnum #1\expandafter \@firstoftwo
 \else \expandafter \@secondoftwo
 \fi
}%
\providecommand \@ifx [1]{%
 \ifx #1\expandafter \@firstoftwo
 \else \expandafter \@secondoftwo
 \fi
}%
\providecommand \natexlab [1]{#1}%
\providecommand \enquote  [1]{``#1''}%
\providecommand \bibnamefont  [1]{#1}%
\providecommand \bibfnamefont [1]{#1}%
\providecommand \citenamefont [1]{#1}%
\providecommand \href@noop [0]{\@secondoftwo}%
\providecommand \href [0]{\begingroup \@sanitize@url \@href}%
\providecommand \@href[1]{\@@startlink{#1}\@@href}%
\providecommand \@@href[1]{\endgroup#1\@@endlink}%
\providecommand \@sanitize@url [0]{\catcode `\\12\catcode `\$12\catcode
  `\&12\catcode `\#12\catcode `\^12\catcode `\_12\catcode `\%12\relax}%
\providecommand \@@startlink[1]{}%
\providecommand \@@endlink[0]{}%
\providecommand \url  [0]{\begingroup\@sanitize@url \@url }%
\providecommand \@url [1]{\endgroup\@href {#1}{\urlprefix }}%
\providecommand \urlprefix  [0]{URL }%
\providecommand \Eprint [0]{\href }%
\providecommand \doibase [0]{http://dx.doi.org/}%
\providecommand \selectlanguage [0]{\@gobble}%
\providecommand \bibinfo  [0]{\@secondoftwo}%
\providecommand \bibfield  [0]{\@secondoftwo}%
\providecommand \translation [1]{[#1]}%
\providecommand \BibitemOpen [0]{}%
\providecommand \bibitemStop [0]{}%
\providecommand \bibitemNoStop [0]{.\EOS\space}%
\providecommand \EOS [0]{\spacefactor3000\relax}%
\providecommand \BibitemShut  [1]{\csname bibitem#1\endcsname}%
\let\auto@bib@innerbib\@empty
\bibitem [{\citenamefont {Yuan}\ \emph {et~al.}(2016)\citenamefont {Yuan},
  \citenamefont {Zhang}, \citenamefont {L\"utkenhaus},\ and\ \citenamefont
  {Ma}}]{YUAN16}%
  \BibitemOpen
  \bibfield  {author} {\bibinfo {author} {\bibfnamefont {X.}~\bibnamefont
  {Yuan}}, \bibinfo {author} {\bibfnamefont {Z.}~\bibnamefont {Zhang}},
  \bibinfo {author} {\bibfnamefont {N.}~\bibnamefont {L\"utkenhaus}}, \ and\
  \bibinfo {author} {\bibfnamefont {X.}~\bibnamefont {Ma}},\ }\href {\doibase
  10.1103/PhysRevA.94.062305} {\bibfield  {journal} {\bibinfo  {journal} {Phys.
  Rev. A}\ }\textbf {\bibinfo {volume} {94}},\ \bibinfo {pages} {062305}
  (\bibinfo {year} {2016})}\BibitemShut {NoStop}%
\bibitem [{\citenamefont {Rahimi-Keshari}\ \emph {et~al.}(2011)\citenamefont
  {Rahimi-Keshari}, \citenamefont {Scherer}, \citenamefont {Mann},
  \citenamefont {Rezakhani}, \citenamefont {Lvovsky},\ and\ \citenamefont
  {Sanders}}]{RAHIMI11}%
  \BibitemOpen
  \bibfield  {author} {\bibinfo {author} {\bibfnamefont {S.}~\bibnamefont
  {Rahimi-Keshari}}, \bibinfo {author} {\bibfnamefont {A.}~\bibnamefont
  {Scherer}}, \bibinfo {author} {\bibfnamefont {A.}~\bibnamefont {Mann}},
  \bibinfo {author} {\bibfnamefont {A.~T.}\ \bibnamefont {Rezakhani}}, \bibinfo
  {author} {\bibfnamefont {A.~I.}\ \bibnamefont {Lvovsky}}, \ and\ \bibinfo
  {author} {\bibfnamefont {B.~C.}\ \bibnamefont {Sanders}},\ }\href
  {http://stacks.iop.org/1367-2630/13/i=1/a=013006} {\bibfield  {journal}
  {\bibinfo  {journal} {New Journal of Physics}\ }\textbf {\bibinfo {volume}
  {13}},\ \bibinfo {pages} {013006} (\bibinfo {year} {2011})}\BibitemShut
  {NoStop}%
\bibitem [{\citenamefont {\ifmmode \check{R}\else
  \v{R}\fi{}eh\'a\ifmmode~\check{c}\else \v{c}\fi{}ek}\ \emph
  {et~al.}(2010)\citenamefont {\ifmmode \check{R}\else
  \v{R}\fi{}eh\'a\ifmmode~\check{c}\else \v{c}\fi{}ek}, \citenamefont
  {Mogilevtsev},\ and\ \citenamefont {Hradil}}]{REHACEK10}%
  \BibitemOpen
  \bibfield  {author} {\bibinfo {author} {\bibfnamefont {J.}~\bibnamefont
  {\ifmmode \check{R}\else \v{R}\fi{}eh\'a\ifmmode~\check{c}\else
  \v{c}\fi{}ek}}, \bibinfo {author} {\bibfnamefont {D.}~\bibnamefont
  {Mogilevtsev}}, \ and\ \bibinfo {author} {\bibfnamefont {Z.}~\bibnamefont
  {Hradil}},\ }\href {\doibase 10.1103/PhysRevLett.105.010402} {\bibfield
  {journal} {\bibinfo  {journal} {Phys. Rev. Lett.}\ }\textbf {\bibinfo
  {volume} {105}},\ \bibinfo {pages} {010402} (\bibinfo {year}
  {2010})}\BibitemShut {NoStop}%
\bibitem [{\citenamefont {Bennett~Ch}\ and\ \citenamefont
  {Brassard}(1984)}]{BENNETT84}%
  \BibitemOpen
  \bibfield  {author} {\bibinfo {author} {\bibfnamefont {H.}~\bibnamefont
  {Bennett~Ch}}\ and\ \bibinfo {author} {\bibfnamefont {G.}~\bibnamefont
  {Brassard}},\ }in\ \href@noop {} {\emph {\bibinfo {booktitle} {Conf. on
  Computers, Systems and Signal Processing (Bangalore, India, Dec. 1984)}}}\
  (\bibinfo {year} {1984})\ pp.\ \bibinfo {pages} {175--9}\BibitemShut
  {NoStop}%
\bibitem [{\citenamefont {Ekert}(1991)}]{EKERT91}%
  \BibitemOpen
  \bibfield  {author} {\bibinfo {author} {\bibfnamefont {A.~K.}\ \bibnamefont
  {Ekert}},\ }\href {\doibase 10.1103/PhysRevLett.67.661} {\bibfield  {journal}
  {\bibinfo  {journal} {Phys. Rev. Lett.}\ }\textbf {\bibinfo {volume} {67}},\
  \bibinfo {pages} {661} (\bibinfo {year} {1991})}\BibitemShut {NoStop}%
\bibitem [{\citenamefont {Schmitt-Manderbach}\ \emph
  {et~al.}(2007)\citenamefont {Schmitt-Manderbach}, \citenamefont {Weier},
  \citenamefont {F{\"u}rst}, \citenamefont {Ursin}, \citenamefont
  {Tiefenbacher}, \citenamefont {Scheidl}, \citenamefont {Perdigues},
  \citenamefont {Sodnik}, \citenamefont {Kurtsiefer}, \citenamefont {Rarity}
  \emph {et~al.}}]{SCHMITT07}%
  \BibitemOpen
  \bibfield  {author} {\bibinfo {author} {\bibfnamefont {T.}~\bibnamefont
  {Schmitt-Manderbach}}, \bibinfo {author} {\bibfnamefont {H.}~\bibnamefont
  {Weier}}, \bibinfo {author} {\bibfnamefont {M.}~\bibnamefont {F{\"u}rst}},
  \bibinfo {author} {\bibfnamefont {R.}~\bibnamefont {Ursin}}, \bibinfo
  {author} {\bibfnamefont {F.}~\bibnamefont {Tiefenbacher}}, \bibinfo {author}
  {\bibfnamefont {T.}~\bibnamefont {Scheidl}}, \bibinfo {author} {\bibfnamefont
  {J.}~\bibnamefont {Perdigues}}, \bibinfo {author} {\bibfnamefont
  {Z.}~\bibnamefont {Sodnik}}, \bibinfo {author} {\bibfnamefont
  {C.}~\bibnamefont {Kurtsiefer}}, \bibinfo {author} {\bibfnamefont {J.~G.}\
  \bibnamefont {Rarity}},  \emph {et~al.},\ }\href@noop {} {\bibfield
  {journal} {\bibinfo  {journal} {Physical Review Letters}\ }\textbf {\bibinfo
  {volume} {98}},\ \bibinfo {pages} {010504} (\bibinfo {year}
  {2007})}\BibitemShut {NoStop}%
\bibitem [{\citenamefont {Lounis}\ and\ \citenamefont
  {Orrit}(2005)}]{LOUNIS05}%
  \BibitemOpen
  \bibfield  {author} {\bibinfo {author} {\bibfnamefont {B.}~\bibnamefont
  {Lounis}}\ and\ \bibinfo {author} {\bibfnamefont {M.}~\bibnamefont {Orrit}},\
  }\href {http://stacks.iop.org/0034-4885/68/i=5/a=R04} {\bibfield  {journal}
  {\bibinfo  {journal} {Reports on Progress in Physics}\ }\textbf {\bibinfo
  {volume} {68}},\ \bibinfo {pages} {1129} (\bibinfo {year}
  {2005})}\BibitemShut {NoStop}%
\bibitem [{\citenamefont {He}\ \emph {et~al.}(2013)\citenamefont {He},
  \citenamefont {He}, \citenamefont {Wei}, \citenamefont {Wu}, \citenamefont
  {Atat{\"u}re}, \citenamefont {Schneider}, \citenamefont {H{\"o}fling},
  \citenamefont {Kamp}, \citenamefont {Lu},\ and\ \citenamefont {Pan}}]{HE13}%
  \BibitemOpen
  \bibfield  {author} {\bibinfo {author} {\bibfnamefont {Y.-M.}\ \bibnamefont
  {He}}, \bibinfo {author} {\bibfnamefont {Y.}~\bibnamefont {He}}, \bibinfo
  {author} {\bibfnamefont {Y.-J.}\ \bibnamefont {Wei}}, \bibinfo {author}
  {\bibfnamefont {D.}~\bibnamefont {Wu}}, \bibinfo {author} {\bibfnamefont
  {M.}~\bibnamefont {Atat{\"u}re}}, \bibinfo {author} {\bibfnamefont
  {C.}~\bibnamefont {Schneider}}, \bibinfo {author} {\bibfnamefont
  {S.}~\bibnamefont {H{\"o}fling}}, \bibinfo {author} {\bibfnamefont
  {M.}~\bibnamefont {Kamp}}, \bibinfo {author} {\bibfnamefont {C.-Y.}\
  \bibnamefont {Lu}}, \ and\ \bibinfo {author} {\bibfnamefont {J.-W.}\
  \bibnamefont {Pan}},\ }\href@noop {} {\bibfield  {journal} {\bibinfo
  {journal} {Nature nanotechnology}\ }\textbf {\bibinfo {volume} {8}},\
  \bibinfo {pages} {213} (\bibinfo {year} {2013})}\BibitemShut {NoStop}%
\bibitem [{\citenamefont {Bennett}\ \emph {et~al.}(1992)\citenamefont
  {Bennett}, \citenamefont {Bessette}, \citenamefont {Brassard}, \citenamefont
  {Salvail},\ and\ \citenamefont {Smolin}}]{BENNETT92}%
  \BibitemOpen
  \bibfield  {author} {\bibinfo {author} {\bibfnamefont {C.~H.}\ \bibnamefont
  {Bennett}}, \bibinfo {author} {\bibfnamefont {F.}~\bibnamefont {Bessette}},
  \bibinfo {author} {\bibfnamefont {G.}~\bibnamefont {Brassard}}, \bibinfo
  {author} {\bibfnamefont {L.}~\bibnamefont {Salvail}}, \ and\ \bibinfo
  {author} {\bibfnamefont {J.}~\bibnamefont {Smolin}},\ }\href {\doibase
  10.1007/BF00191318} {\bibfield  {journal} {\bibinfo  {journal} {Journal of
  Cryptology}\ }\textbf {\bibinfo {volume} {5}},\ \bibinfo {pages} {3}
  (\bibinfo {year} {1992})}\BibitemShut {NoStop}%
\bibitem [{\citenamefont {Du\ifmmode~\check{s}\else \v{s}\fi{}ek}\ \emph
  {et~al.}(2000)\citenamefont {Du\ifmmode~\check{s}\else \v{s}\fi{}ek},
  \citenamefont {Jahma},\ and\ \citenamefont {L\"utkenhaus}}]{DUSEK00}%
  \BibitemOpen
  \bibfield  {author} {\bibinfo {author} {\bibfnamefont {M.}~\bibnamefont
  {Du\ifmmode~\check{s}\else \v{s}\fi{}ek}}, \bibinfo {author} {\bibfnamefont
  {M.}~\bibnamefont {Jahma}}, \ and\ \bibinfo {author} {\bibfnamefont
  {N.}~\bibnamefont {L\"utkenhaus}},\ }\href {\doibase
  10.1103/PhysRevA.62.022306} {\bibfield  {journal} {\bibinfo  {journal} {Phys.
  Rev. A}\ }\textbf {\bibinfo {volume} {62}},\ \bibinfo {pages} {022306}
  (\bibinfo {year} {2000})}\BibitemShut {NoStop}%
\bibitem [{\citenamefont {Brassard}\ \emph {et~al.}(2000)\citenamefont
  {Brassard}, \citenamefont {L\"utkenhaus}, \citenamefont {Mor},\ and\
  \citenamefont {Sanders}}]{BRASSARD00}%
  \BibitemOpen
  \bibfield  {author} {\bibinfo {author} {\bibfnamefont {G.}~\bibnamefont
  {Brassard}}, \bibinfo {author} {\bibfnamefont {N.}~\bibnamefont
  {L\"utkenhaus}}, \bibinfo {author} {\bibfnamefont {T.}~\bibnamefont {Mor}}, \
  and\ \bibinfo {author} {\bibfnamefont {B.~C.}\ \bibnamefont {Sanders}},\
  }\href {\doibase 10.1103/PhysRevLett.85.1330} {\bibfield  {journal} {\bibinfo
   {journal} {Phys. Rev. Lett.}\ }\textbf {\bibinfo {volume} {85}},\ \bibinfo
  {pages} {1330} (\bibinfo {year} {2000})}\BibitemShut {NoStop}%
\bibitem [{\citenamefont {Hwang}(2003)}]{HWANG03}%
  \BibitemOpen
  \bibfield  {author} {\bibinfo {author} {\bibfnamefont {W.-Y.}\ \bibnamefont
  {Hwang}},\ }\href {\doibase 10.1103/PhysRevLett.91.057901} {\bibfield
  {journal} {\bibinfo  {journal} {Phys. Rev. Lett.}\ }\textbf {\bibinfo
  {volume} {91}},\ \bibinfo {pages} {057901} (\bibinfo {year}
  {2003})}\BibitemShut {NoStop}%
\bibitem [{\citenamefont {Gottesman}\ \emph {et~al.}(2004)\citenamefont
  {Gottesman}, \citenamefont {Lo}, \citenamefont {Lutkenhaus},\ and\
  \citenamefont {Preskill}}]{GOTTESMAN04}%
  \BibitemOpen
  \bibfield  {author} {\bibinfo {author} {\bibfnamefont {D.}~\bibnamefont
  {Gottesman}}, \bibinfo {author} {\bibfnamefont {H.~K.}\ \bibnamefont {Lo}},
  \bibinfo {author} {\bibfnamefont {N.}~\bibnamefont {Lutkenhaus}}, \ and\
  \bibinfo {author} {\bibfnamefont {J.}~\bibnamefont {Preskill}},\ }in\ \href
  {\doibase 10.1109/ISIT.2004.1365172} {\emph {\bibinfo {booktitle}
  {Information Theory, 2004. ISIT 2004. Proceedings. International Symposium
  on}}}\ (\bibinfo {year} {2004})\ pp.\ \bibinfo {pages} {136--}\BibitemShut
  {NoStop}%
\bibitem [{\citenamefont {Lo}\ \emph {et~al.}(2005)\citenamefont {Lo},
  \citenamefont {Ma},\ and\ \citenamefont {Chen}}]{LO05}%
  \BibitemOpen
  \bibfield  {author} {\bibinfo {author} {\bibfnamefont {H.-K.}\ \bibnamefont
  {Lo}}, \bibinfo {author} {\bibfnamefont {X.}~\bibnamefont {Ma}}, \ and\
  \bibinfo {author} {\bibfnamefont {K.}~\bibnamefont {Chen}},\ }\href@noop {}
  {\bibfield  {journal} {\bibinfo  {journal} {Physical review letters}\
  }\textbf {\bibinfo {volume} {94}},\ \bibinfo {pages} {230504} (\bibinfo
  {year} {2005})}\BibitemShut {NoStop}%
\bibitem [{\citenamefont {Ma}\ \emph {et~al.}(2005)\citenamefont {Ma},
  \citenamefont {Qi}, \citenamefont {Zhao},\ and\ \citenamefont {Lo}}]{MA05}%
  \BibitemOpen
  \bibfield  {author} {\bibinfo {author} {\bibfnamefont {X.}~\bibnamefont
  {Ma}}, \bibinfo {author} {\bibfnamefont {B.}~\bibnamefont {Qi}}, \bibinfo
  {author} {\bibfnamefont {Y.}~\bibnamefont {Zhao}}, \ and\ \bibinfo {author}
  {\bibfnamefont {H.-K.}\ \bibnamefont {Lo}},\ }\href@noop {} {\bibfield
  {journal} {\bibinfo  {journal} {Physical Review A}\ }\textbf {\bibinfo
  {volume} {72}},\ \bibinfo {pages} {012326} (\bibinfo {year}
  {2005})}\BibitemShut {NoStop}%
\bibitem [{\citenamefont {Wang}(2005)}]{WANG05}%
  \BibitemOpen
  \bibfield  {author} {\bibinfo {author} {\bibfnamefont {X.-B.}\ \bibnamefont
  {Wang}},\ }\href {\doibase 10.1103/PhysRevLett.94.230503} {\bibfield
  {journal} {\bibinfo  {journal} {Phys. Rev. Lett.}\ }\textbf {\bibinfo
  {volume} {94}},\ \bibinfo {pages} {230503} (\bibinfo {year}
  {2005})}\BibitemShut {NoStop}%
\bibitem [{\citenamefont {Zhao}\ \emph {et~al.}(2006)\citenamefont {Zhao},
  \citenamefont {Qi}, \citenamefont {Ma}, \citenamefont {Lo},\ and\
  \citenamefont {Qian}}]{ZHAO06}%
  \BibitemOpen
  \bibfield  {author} {\bibinfo {author} {\bibfnamefont {Y.}~\bibnamefont
  {Zhao}}, \bibinfo {author} {\bibfnamefont {B.}~\bibnamefont {Qi}}, \bibinfo
  {author} {\bibfnamefont {X.}~\bibnamefont {Ma}}, \bibinfo {author}
  {\bibfnamefont {H.-K.}\ \bibnamefont {Lo}}, \ and\ \bibinfo {author}
  {\bibfnamefont {L.}~\bibnamefont {Qian}},\ }\href {\doibase
  10.1103/PhysRevLett.96.070502} {\bibfield  {journal} {\bibinfo  {journal}
  {Phys. Rev. Lett.}\ }\textbf {\bibinfo {volume} {96}},\ \bibinfo {pages}
  {070502} (\bibinfo {year} {2006})}\BibitemShut {NoStop}%
\bibitem [{\citenamefont {Inamori}\ \emph {et~al.}(2007)\citenamefont
  {Inamori}, \citenamefont {L{\"u}tkenhaus},\ and\ \citenamefont
  {Mayers}}]{INAMORI07}%
  \BibitemOpen
  \bibfield  {author} {\bibinfo {author} {\bibfnamefont {H.}~\bibnamefont
  {Inamori}}, \bibinfo {author} {\bibfnamefont {N.}~\bibnamefont
  {L{\"u}tkenhaus}}, \ and\ \bibinfo {author} {\bibfnamefont {D.}~\bibnamefont
  {Mayers}},\ }\href {\doibase 10.1140/epjd/e2007-00010-4} {\bibfield
  {journal} {\bibinfo  {journal} {The European Physical Journal D}\ }\textbf
  {\bibinfo {volume} {41}},\ \bibinfo {pages} {599} (\bibinfo {year}
  {2007})}\BibitemShut {NoStop}%
\bibitem [{\citenamefont {Sun}\ \emph {et~al.}(2013)\citenamefont {Sun},
  \citenamefont {Gao}, \citenamefont {Li},\ and\ \citenamefont
  {Liang}}]{SUN13}%
  \BibitemOpen
  \bibfield  {author} {\bibinfo {author} {\bibfnamefont {S.-H.}\ \bibnamefont
  {Sun}}, \bibinfo {author} {\bibfnamefont {M.}~\bibnamefont {Gao}}, \bibinfo
  {author} {\bibfnamefont {C.-Y.}\ \bibnamefont {Li}}, \ and\ \bibinfo {author}
  {\bibfnamefont {L.-M.}\ \bibnamefont {Liang}},\ }\href {\doibase
  10.1103/PhysRevA.87.052329} {\bibfield  {journal} {\bibinfo  {journal} {Phys.
  Rev. A}\ }\textbf {\bibinfo {volume} {87}},\ \bibinfo {pages} {052329}
  (\bibinfo {year} {2013})}\BibitemShut {NoStop}%
\bibitem [{\citenamefont {Lim}\ \emph {et~al.}(2014)\citenamefont {Lim},
  \citenamefont {Curty}, \citenamefont {Walenta}, \citenamefont {Xu},\ and\
  \citenamefont {Zbinden}}]{LIM14}%
  \BibitemOpen
  \bibfield  {author} {\bibinfo {author} {\bibfnamefont {C.~C.~W.}\
  \bibnamefont {Lim}}, \bibinfo {author} {\bibfnamefont {M.}~\bibnamefont
  {Curty}}, \bibinfo {author} {\bibfnamefont {N.}~\bibnamefont {Walenta}},
  \bibinfo {author} {\bibfnamefont {F.}~\bibnamefont {Xu}}, \ and\ \bibinfo
  {author} {\bibfnamefont {H.}~\bibnamefont {Zbinden}},\ }\href {\doibase
  10.1103/PhysRevA.89.022307} {\bibfield  {journal} {\bibinfo  {journal} {Phys.
  Rev. A}\ }\textbf {\bibinfo {volume} {89}},\ \bibinfo {pages} {022307}
  (\bibinfo {year} {2014})}\BibitemShut {NoStop}%
\bibitem [{\citenamefont {Wang}\ and\ \citenamefont {Wang}(2014)}]{WANG14}%
  \BibitemOpen
  \bibfield  {author} {\bibinfo {author} {\bibfnamefont {Q.}~\bibnamefont
  {Wang}}\ and\ \bibinfo {author} {\bibfnamefont {X.-B.}\ \bibnamefont
  {Wang}},\ }\href@noop {} {\bibfield  {journal} {\bibinfo  {journal}
  {Scientific reports}\ }\textbf {\bibinfo {volume} {4}} (\bibinfo {year}
  {2014})}\BibitemShut {NoStop}%
\bibitem [{\citenamefont {Lvovsky}\ \emph {et~al.}(2001)\citenamefont
  {Lvovsky}, \citenamefont {Hansen}, \citenamefont {Aichele}, \citenamefont
  {Benson}, \citenamefont {Mlynek},\ and\ \citenamefont
  {Schiller}}]{LVOVSKY01}%
  \BibitemOpen
  \bibfield  {author} {\bibinfo {author} {\bibfnamefont {A.~I.}\ \bibnamefont
  {Lvovsky}}, \bibinfo {author} {\bibfnamefont {H.}~\bibnamefont {Hansen}},
  \bibinfo {author} {\bibfnamefont {T.}~\bibnamefont {Aichele}}, \bibinfo
  {author} {\bibfnamefont {O.}~\bibnamefont {Benson}}, \bibinfo {author}
  {\bibfnamefont {J.}~\bibnamefont {Mlynek}}, \ and\ \bibinfo {author}
  {\bibfnamefont {S.}~\bibnamefont {Schiller}},\ }\href@noop {} {\bibfield
  {journal} {\bibinfo  {journal} {Physical Review Letters}\ }\textbf {\bibinfo
  {volume} {87}},\ \bibinfo {pages} {050402} (\bibinfo {year}
  {2001})}\BibitemShut {NoStop}%
\bibitem [{Com()}]{Comment}%
  \BibitemOpen
  \href@noop {} {}\bibinfo {note} {Our result could be seen as a particular
  case of the operational tomography method suggested in \cite{REHACEK10} using
  PRCS as reference states. However, our method is not based in a minimization
  procedure. Instead, the state reconstructions makes explicit use of the
  relationship between PRCS and Fock states (Eqs. \ref{mat_densidad} and
  \ref{Pk})}\BibitemShut {NoStop}%
\bibitem [{\citenamefont {Lvovsky}\ and\ \citenamefont
  {Raymer}(2009)}]{LVOVSKY09}%
  \BibitemOpen
  \bibfield  {author} {\bibinfo {author} {\bibfnamefont {A.~I.}\ \bibnamefont
  {Lvovsky}}\ and\ \bibinfo {author} {\bibfnamefont {M.~G.}\ \bibnamefont
  {Raymer}},\ }\href@noop {} {\bibfield  {journal} {\bibinfo  {journal}
  {Reviews of Modern Physics}\ }\textbf {\bibinfo {volume} {81}},\ \bibinfo
  {pages} {299} (\bibinfo {year} {2009})}\BibitemShut {NoStop}%
\bibitem [{\citenamefont {Munroe}\ \emph {et~al.}(1995)\citenamefont {Munroe},
  \citenamefont {Boggavarapu}, \citenamefont {Anderson},\ and\ \citenamefont
  {Raymer}}]{MUNROE95}%
  \BibitemOpen
  \bibfield  {author} {\bibinfo {author} {\bibfnamefont {M.}~\bibnamefont
  {Munroe}}, \bibinfo {author} {\bibfnamefont {D.}~\bibnamefont {Boggavarapu}},
  \bibinfo {author} {\bibfnamefont {M.~E.}\ \bibnamefont {Anderson}}, \ and\
  \bibinfo {author} {\bibfnamefont {M.~G.}\ \bibnamefont {Raymer}},\ }\href
  {\doibase 10.1103/PhysRevA.52.R924} {\bibfield  {journal} {\bibinfo
  {journal} {Phys. Rev. A}\ }\textbf {\bibinfo {volume} {52}},\ \bibinfo
  {pages} {R924} (\bibinfo {year} {1995})}\BibitemShut {NoStop}%
\bibitem [{\citenamefont {Gerry}\ and\ \citenamefont {Knight}(2005)}]{GERRY05}%
  \BibitemOpen
  \bibfield  {author} {\bibinfo {author} {\bibfnamefont {C.}~\bibnamefont
  {Gerry}}\ and\ \bibinfo {author} {\bibfnamefont {P.}~\bibnamefont {Knight}},\
  }\href@noop {} {\emph {\bibinfo {title} {Introductory quantum optics}}}\
  (\bibinfo  {publisher} {Cambridge university press},\ \bibinfo {year}
  {2005})\BibitemShut {NoStop}%
\end{thebibliography}

%

\end{document}